\documentclass{ws-ijmpe}
\usepackage[super,compress]{cite}
\usepackage{lineno}
\begin{document}

\markboth{J. Singh, M.~U.~Ashraf, A.~M.~Khan, A.~Rana, and S.~Kabana}{Strange hadron production in $O+O$ collisions}

\catchline{}{}{}{}{}

\title{Anisotropic flow coefficients for charged hadrons in $O+O$ collisions at $\sqrt{s_{\mathrm{NN}}}~=~7$ TeV using AMPT}

\author{J.~Singh\footnote{J.~Singh}}

\address{Instituto de Alta Investigaci\'on, Universidad de Tarapac\'a, Casilla 7D, Arica, 1000000, Chile\\ 
jsingh2@bnl.gov}

\author{M.~U.~Ashraf}
\address{Department of Physics and Astronomy, Wayne State University, 666 W. Hancock, Detroit, Michigan 48201, USA\\
}

\author{A.~M.~Khan}
\address{Instituto de Alta Investigaci\'on, Universidad de Tarapac\'a, Casilla 7D, Arica, 1000000, Chile\\
}

\author{A.~Rana}
\address{Department of Physics, D.A.V. College, Sector 10, Chandigarh 160011, India\\
}

\author{S.~Kabana}
\address{Instituto de Alta Investigaci\'on, Universidad de Tarapac\'a, Casilla 7D, Arica, 1000000, Chile\\
}

\maketitle
\begin{history}
\received{Day Month Year}
\end{history}

\begin{abstract}
In this article, we report on the predictions of $v_2$, $v_3$ and $v_4$ for charged hadrons in O+O collisions at $\sqrt{s_{\mathrm{NN}}}~=~7$~TeV using both AMPT-default and AMPT-String Melting. These predictions are compared with the existing published data of $p+p$, $p+Pb$, and $Pb+Pb$ collisions at LHC energies. The transverse momentum ($p_T$) dependence of $v_2$ is also investigated for different centrality classes. $O+O$ collisions provide a unique opportunity to bridge the gap between small and large collision systems, offering critical insight into the onset of collective behavior in QCD matter. 
\end{abstract}


\ccode{PACS numbers: 25.75.-q, 25.75.Ld}

\section{Introduction}

The primary objective of the heavy-ion physics programs at the Relativistic Heavy Ion Collider (RHIC) and the Large Hadron Collider (LHC) is to explore the properties of nuclear matter under extreme conditions, particularly the formation and characteristics of the quark–gluon plasma (QGP)\cite{qgp1,qgp2,qgp3}. Initial results from RHIC experiments~\cite{brahms1,phobos1,star1,phenix1} provided compelling evidence that the QGP exhibits collective behavior characteristic of a nearly perfect fluid, characterized by strong interactions and a remarkably low shear viscosity to entropy density ratio ($\eta/s$)\cite{th1,th2}.

A key observable in probing the QGP is anisotropic flow, which originates from pressure gradients established by the initial spatial anisotropy in a collision and observed as a momentum anisotropy in the final-state particle distribution. The anisotropic flow of the system is typically characterized by flow harmonics, $v_n$, which are the Fourier coefficients of the azimuthal distribution of particles, $v_{n} = \langle cos [n(\phi - \Psi_{RP})] \rangle$~\cite{fourier_flow}. Here, $n$ is the order of the flow harmonic, $\phi$ is the azimuthal angle and $\Psi_{RP}$ is the reaction plane angle of harmonic $n$.  The first four harmonics, $v_1$, $v_2$, $v_3$ and $v_4$, are referred to as directed, elliptic, triangular, and quadrangular (or hexadecapole) flow, respectively. Extensive measurements of these harmonics have been carried out at the RHIC~\cite{rhicflow1,rhicflow2,rhicflow3} and the LHC~\cite{lhcflow1,lhcflow2,lhcflow3,lhcflow4}. The observed anisotropic flow provides compelling evidence for collective behavior in the system~\cite{anth1} and is successfully described by relativistic hydrodynamic models, indicating the formation of a near-perfect fluid state~\cite{anth2}.

While traditional studies of the QGP have primarily focused on large systems such as $Au+Au$, $Pb+Pb$ and $Xe+Xe$ collisions, recent measurements in small systems (e.g., $pp$ and $p+Pb$) have revealed signatures reminiscent of collectivity, raising important questions about the onset of QGP-like behavior in smaller collision systems~\cite{small1,small2,small3}. Recent theoretical advancements and experimental observations suggest a dominance of final-state effects in high-multiplicity $pp$ and $p+Pb$ collisions, potentially leading to the formation of QGP droplets in these small systems. Conversely, for low-multiplicity collisions, initial-state momentum correlations emerging from gluon saturation are postulated to play a significant role~\cite{gluoncorr}.

In this context, oxygen-oxygen ($O+O$) collisions at $\sqrt{s_{\rm NN}}~=~7$~TeV serve as an important intermediate system at the LHC. With a system size between that of $pp$ and $Pb+Pb$ collisions, $O+O$ collisions offer a unique opportunity to investigate the emergence of collective phenomena and constrain the system-size dependence of QGP-like behavior~\cite{OO1, OO2, smallsystem1}. LHC is anticipated to take data from $O+O$ collisions and the study of this system holds immense potential to significantly enhance our understanding of this phenomena. 
The underlying mechanisms responsible for particle production in $O+O$ collisions have attracted significant theoretical attention in recent studies~\cite{ooth1,ooth2}.


The main purpose of this paper is to study the flow harmonics $v_n$ as a function of produced particle multiplicity ($N_{ch}$) and transverse momentum ($p_T$) in $O+O$ collisions at $\sqrt{s_{\mathrm{NN}}}~=~7$~TeV with both versions of AMPT~\cite{AMPT1} model. We selected AMPT due to its demonstrated success in describing anisotropic flow across a range of collision system sizes, from large to small~\cite{AMPT2,AMPT3,AMPT4}.

\section{Methodology}
In this section, we describe the event generators employed in this study and the analysis techniques used to extract flow harmonics.

\subsection{AMPT}
A Multi-Phase Transport (AMPT) model is a hybrid transport model developed to investigate the dynamics of relativistic heavy-ion collisions and has been widely employed to study various observables at both RHIC and LHC energies~\cite{AMPT1,AMPT2}. In the AMPT model, the initial conditions, including spatial and momentum distributions of minijet partons and soft string excitations—are generated using the HIJING model~\cite{hijing1}. The subsequent space-time evolution of partons is governed by the Zhang’s Parton Cascade (ZPC) model~\cite{cascade1}. After the partonic phase, the remaining degrees of freedom are converted into hadrons either through Lund string fragmentation (in the default version) or via quark coalescence (in the string melting version). Final-state hadronic interactions are modeled using the A Relativistic Transport (ART) model~\cite{art1}. The default version (AMPT-Def), which includes only minijet partons in the parton cascade and hadronizes via string fragmentation, reproduces basic observables such as rapidity distributions and transverse momentum spectra of identified particles at SPS and RHIC energies~\cite{AMPT3}. The string melting version (AMPT-SM), on the other hand, converts all excited strings into their valence quarks, which then undergo scatterings and recombine via a simple quark coalescence model. This approach has been successful in describing anisotropic flow signals in both large and small systems~\cite{AMPT3,AMPT4}.

In this study, both AMPT-Def and AMPT-SM versions are used to simulate $\sim$6 million minimum-bias events each for $O+O$ collisions at $\sqrt{s_{\mathrm{NN}}}~=~7$~TeV. The centrality classes are defined based on the pseudorapidity distribution of charged particles within $\rvert \Delta\eta \rvert>0.5$, consistent with the methodology adopted by the LHC experiments. The corresponding values of $\langle dN_{\text{ch}}/d\eta \rangle$ and $\langle N_{\text{part}} \rangle$ for different centrality classes in both versions of the model are provided in Refs.~\cite{Identified_OO, strange_OO7}.

\subsection{Analysis technique}
To quantify anisotropic flow in $O+O$ collisions, we employ the two-particle correlation method using the Q-cumulant formalism~\cite{Qcumulant}. Direct computation of multi-particle correlations is computationally intensive, and the Q-cumulant method provides an efficient and numerically stable alternative~\cite{Qcumulant}. 
The Q-vector for harmonic $n$ is defined as:
\begin{equation}
  Q_n = \sum_{i=1}^M e^{in\phi_i}
  \label{QVector}
\end{equation}
where $n$ represents the harmonic order (e.g., $n = 1,~2,~3,~...$), $\phi_i$ is the azimuthal angle of the $i^{th}$ particle and the summation is performed over all particles in the event with a multiplicity of $M$. The computation of two-particle correlations is performed in two steps:\\
{\bf{Step-I}}: The following expression can be used to define the average two-particle correlations for each event:
\begin{equation}
  \langle 2 \rangle_{n\mid n} \equiv \frac{1}{P(M,2)} \sum_{\substack{i,j=1 (i\neq j)}}^M e^{in(\phi_{i}-\phi_{j})}
  \label{qneq2}
\end{equation}

where $P(n,m)$ is $\frac{n!}{(n-m)!}$\\
{\bf{Step-II}}: After determining the correlations for each event, the final two-particle azimuthal correlation is determined by taking an average across all events:
\begin{equation}
  \langle \langle 2 \rangle \rangle_{n|n} \equiv \frac{\sum_{i=1}^N (w_2)_i (\langle 2 \rangle_{n|n})_i}  {\sum_{i=1}^N (w_2)_i}
  \label{qneq22}
\end{equation}
where double brackets denote an average over all tracks and then over all ``N'' events. The multiplicity weight ($w_2$) is introduced to get rid of the multiplicity (M) fluctuations which are defined as, $w_2 \equiv M(M-1)$. Thus the cumulant is defined as:
\begin{equation}
c_2\{ 2 \} = \langle \langle 2 \rangle \rangle
\label{cn2eq}
\end{equation}
In terms of flow harmonic, the cumulants can be written as,
\begin{equation}
v_n^2\{2\} = c_2\{ 2 \}
\label{v2c2eq}
\end{equation}
This method allows for the precise extraction of flow coefficients $v_n$ while reducing contributions from non-flow effects, such as resonance decays and jets, particularly by applying a pseudorapidity gap ($\rvert \Delta\eta \rvert>0.5$) between correlated particles.

\section{Results and Discussions}
\begin{figure}[th]
    \centering       
    \includegraphics[width=0.80\linewidth]{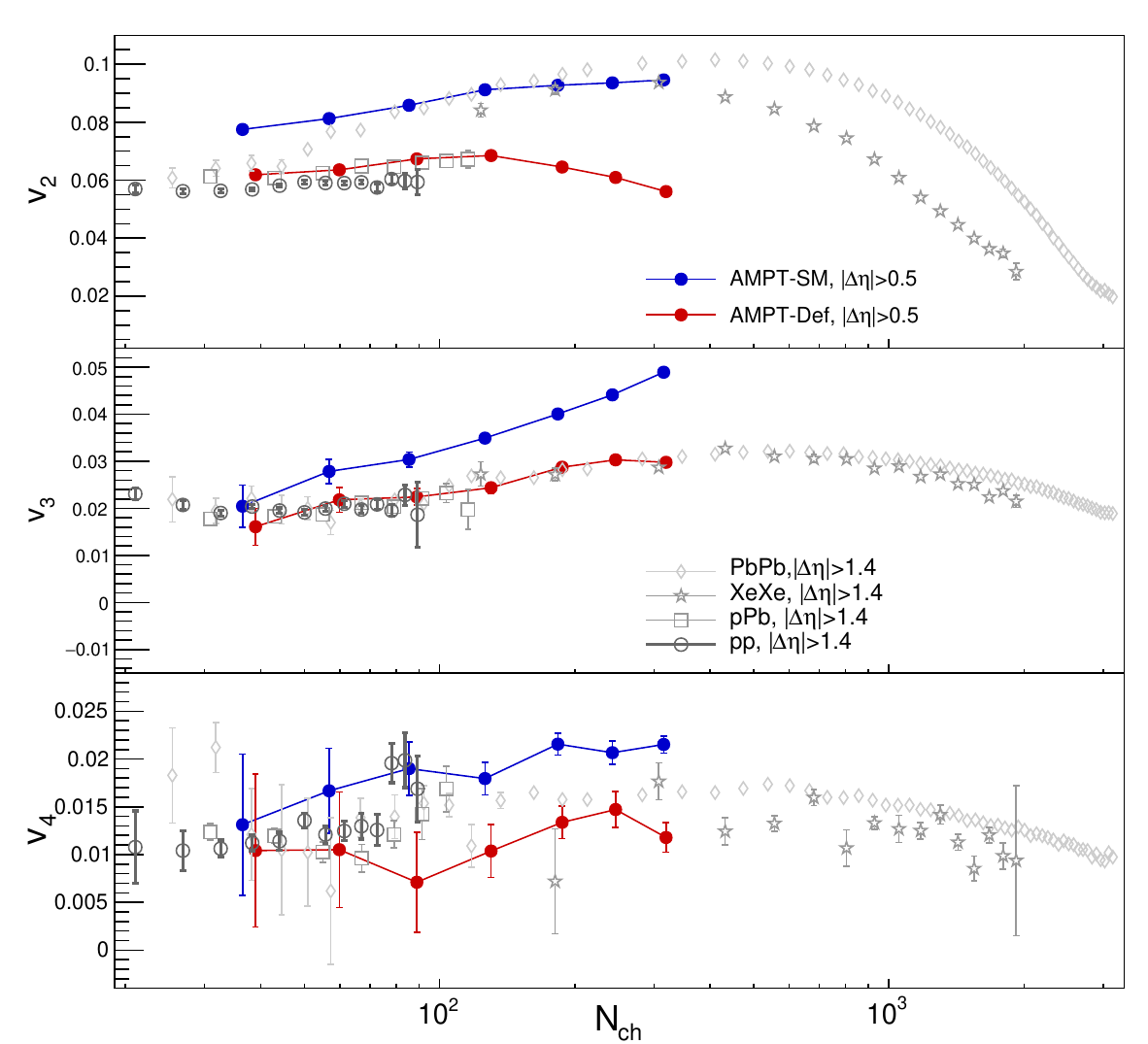} 
    \caption{(Color online) Multiplicity dependence of $v_n$ for $O+O$ collisions at $\sqrt{s_{\mathrm{NN}}}=7$ TeV, using AMPT-SM and AMPT-Def. The results are compared with experimental data from $p+p$, $p+Pb$, $Xe+Xe$, and $Pb+Pb$ collisions at various energies from the LHC~\cite{ALICE_flow}.}\label{fig:flow_vs_Nch}
\end{figure}

Figure~\ref{fig:flow_vs_Nch} presents the anisotropic flow coefficients $v_2$, $v_3$, and $v_4$ as a function of charged particle multiplicity ($N_{ch}$) for O+O collisions at $\sqrt{s_{NN}}=7$ TeV, predicted using the AMPT-SM and AMPT-Def. 
The results are further compared with experimental data from $p+p$, $p+Pb$, $Xe+Xe$, and $Pb+Pb$ collisions at various LHC energies~\cite{ALICE_flow}.
\begin{figure}[th]
    \centering       
    \includegraphics[width=0.49\linewidth]{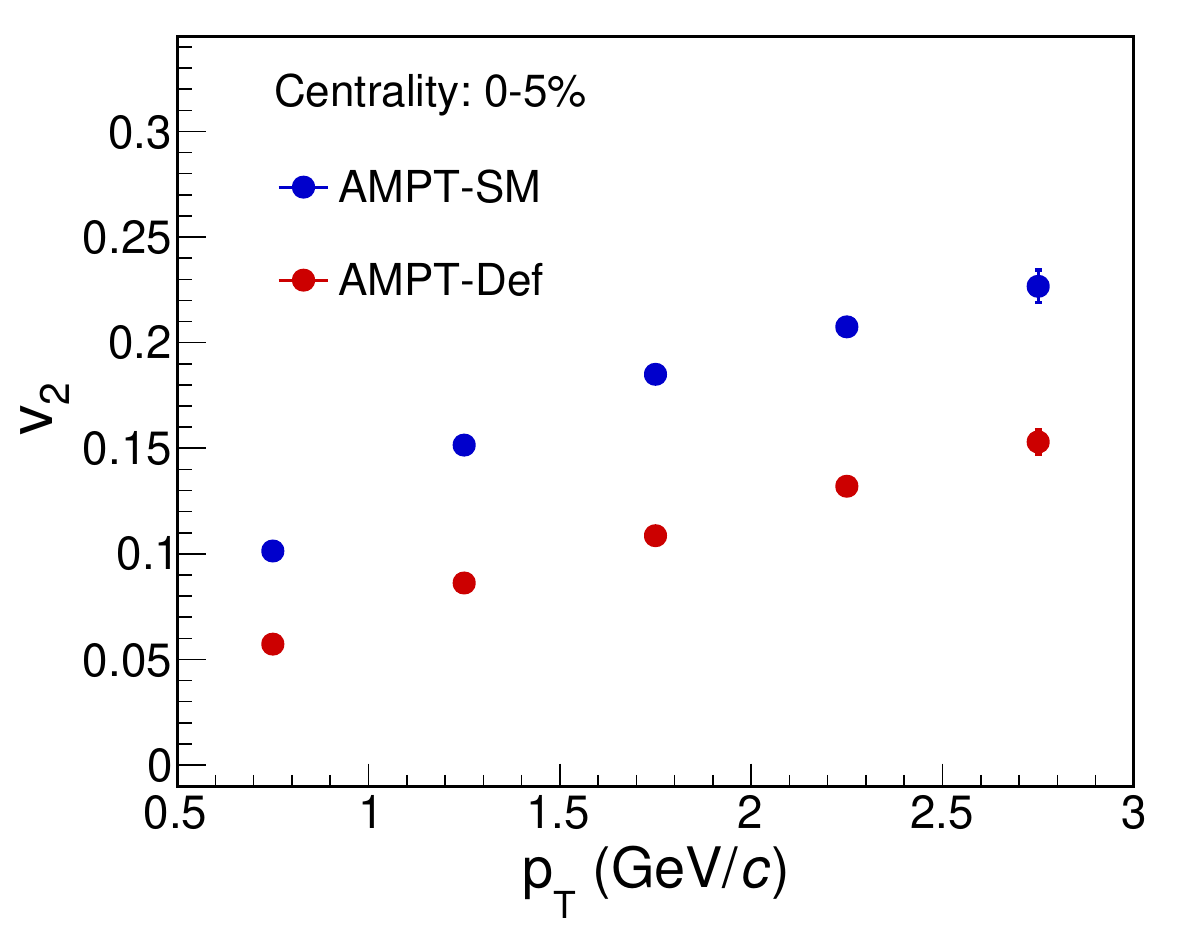} 
    \includegraphics[width=0.49\linewidth]{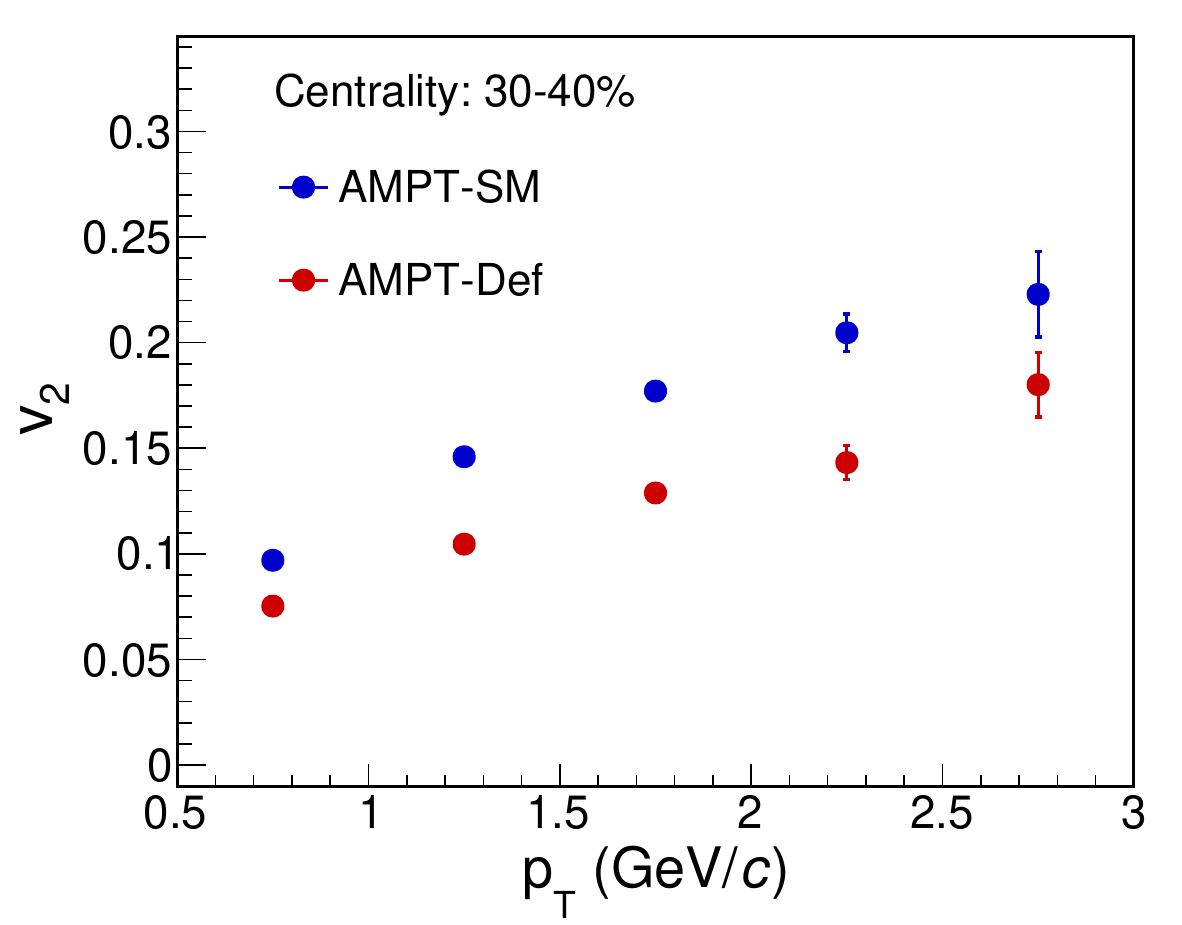}
    \caption{(Color online) $v_2$ for all charged hadrons versus transverse momentum ($p_T$) in $O+O$ collisions at $\sqrt{s_{\mathrm{NN}}}=7$ TeV, shown for 0-5$\%$ (left) and 30-40$\%$ (right) collision centrality for AMPT-SM and AMPT-Def.}\label{fig:v2pt}
\end{figure}
It is observed that $v_n$ coefficients exhibit weak dependence on multiplicity for $O+O$ collisions. Notably, the predicted $v_n$ values for $O+O$ collisions exhibit significant overlap in final-state multiplicity with those measured in both small (e.g., $p+p$, $p+Pb$) and large (e.g., $Xe+Xe$, $Pb+Pb$) collision systems. This overlap reinforces the idea that collective behavior, typically associated with the formation of the QGP, may also emerge in smaller systems given sufficient final-state particle density. The results suggest the possible development of a strongly interacting medium in $O+O$ collisions, motivating future experimental investigation during the upcoming LHC $O+O$ run.

Figure~\ref{fig:v2pt} displays the elliptic flow coefficient ($v_2$) of all charged hadrons as a function of transverse momentum ($p_T$) in $O+O$ collisions at $\sqrt{s_{\mathrm{NN}}}=7$~TeV, shown separately for the 0–5$\%$ (left) and 30–40$\%$ (right) centrality classes using AMPT model. In both centralities, $v_2$ increases monotonically with $p_T$, reflecting the buildup of momentum anisotropy during the system's collective expansion. The AMPT-SM model systematically predicts higher $v_2$ values compared to AMPT-Def, indicating a stronger partonic interaction and enhanced collectivity due to the string melting mechanism. This rising trend with $p_T$ underscores the important role of early-stage partonic dynamics and subsequent hydrodynamic evolution in the development of azimuthal anisotropy. These results are part of an ongoing study and are subject to refinement through continued analysis and model optimization.


\section{Conclusions}
We present the predictions for anisotropic flow coefficients ($v_n$, with $n = 2, 3, 4$) in $O+O$ collisions at $\sqrt{s_{\mathrm{NN}}} = 7$ TeV using AMPT-Default and AMPT-String Melting. The results from AMPT-Def shows good agreement with existing experimental measurements from $p+p$, $p+Pb$, $Xe+Xe$, and $Pb+Pb$ collisions across various LHC energies. A notable overlap in final-state charged-particle multiplicity is observed between $O+O$ collisions and both small and large collision systems, reinforcing the relevance of $O+O$ as an intermediate-size system for investigating collective behavior. The $p_T$-differential $v_2$ exhibits a consistent increasing trend, with AMPT-SM producing systematically larger values due to enhanced partonic interactions. These results suggest the emergence of strong final-state effects and potential QGP-like collectivity even in light-ion collisions. 
The upcoming LHC data on $O+O$ collisions will be crucial for constraining theoretical models and improving our understanding of the system-size dependence of QGP-like phenomena.

\end{document}